\documentclass[12pt]{article}
\begin{document}
\title{Comment on Faizal et al EPJC 76:30}
\author{B.G. Sidharth\\
G.P. Birla Observatory \& Astronomical Research Centre\\
B.M. Birla Science Centre, Adarsh Nagar, Hyderabad - 500 063
(India)}
\date{}
\maketitle
\begin{abstract}
In a recent paper in EPJC January 2016, Faizal, Khalil and Das have
proposed time crystals with duration several orders of magnitude
greater than Planck scale. We comment on this paper and shed further
light on this aspect.
\end{abstract}
Recently Faizal et al., in this journal \cite{epjc} described time
crystals which are several orders of magnitude greater than the
Planck scale. The discreteness and hence noncommutativity of
spacetime has been considered in the literature from the 1940s. In
the earlier attempts the scale at which this discreteness takes
place has been the Planck scale. Several authors like Snyder,
Schild, Kadyshevskii, Ginsburg, Caldirola and others have considered
this discreteness, as also in very recent Quantum Gravity approaches
\cite{bgscu,tduniv}. However the author considered discreteness at
the Compton Scale to develop his successful cosmology of 1997
\cite{bgsijmpa,bgsijtp,bgsmg8}. This predicted in advance a slowly
accelerating universe driven by what we today call dark energy, when
the standard big bang model said exactly the opposite. It was of
course argued at length by Wigner and Salecker \cite{wigner} in the
late fifties that there cannot be a physical time within the Compton
Scale. Further the author showed more than 12 years ago in several
papers in Foundation of Physics and Chaos, Solitons and Fractals,
how the coherent Compton Scale arises from the Planck Scale through
a coherence approach including the Landau-Ginsburg phase transition
\cite{bgsfpl,bgscsf}. So, even though as in the Prigogine cosmology
a Big Bang event would lead to the Planck scale or Wheeler's Quantum
Foam \cite{wheeler}, this would lead to a several order of magnitude
higher scale through phase transition. In fact just prior to the
phase transition we would have
\begin{equation}
- \frac{\hbar^2}{2m} \nabla^2 \psi + \beta |\psi |^2 \psi = - \alpha
\psi\label{3eea7}
\end{equation}
In (\ref{3eea7}) $\psi$ denotes the wave function of the particle at
a point which is in the impenetrable Planck length. Its derivation
is explained in \cite{tduniv,bgscsf}-- but basically it stems from a
simple two or more state model of probability amplitudes first
worked out by Feynman.\\
Equation (\ref{3eea7}) leads to the Landau-Ginsberg phase transition
with coherence length
\begin{equation}
\xi = \left(\frac{\gamma}{\alpha}\right)^{\frac{1}{2}}\label{3eea9}
\end{equation}
$\xi$ which is in the left side is the coherence length, $\gamma$ is
$\hbar^2 / 2m$ is in the landau theory and $\alpha = mc^2$ is the
energy.\\
This is the Compton scale (Cf.ref.\cite{tduniv}) in our case.\\
More recently this was also shown by Beck and Murray \cite{beck} and
even more recently it was argued in The European Physical Journal C by Faizal, Khalil and Das \cite{epjc}.\\
On the contrary sticking to the Planck Scale without such a phase
transition could prove disastrous as recently articulated by Harry
Cliff of Cambridge University and the LHC Collaboration - it would
lead to the end of physics, particularly because of the cosmological
constant being, in this case $10^{120}$ times its observed value
\cite{weinberg,harry}.\\ \\
{\large {\bf Acknowledgment}}\\ \\
I am thankful for the pertinent points raised by the referees.

\noindent {\large{\bf APPENDIX (OPTIONAL)}}\\ \\
1. Let us see this in a little greater detail: Starting with a
simple superposition of states model, first invoked by Feynman, we
have:
$$
\psi_\imath (t - \Delta t) - \psi_\imath (t + \Delta t) = \sum_{j}
\left[\delta_{\imath j} - \frac{\imath}{\hbar} H_{\imath
j}(t)\right] \psi_\imath (t)\quad \quad \quad (1)$$

In the limit this can be shown to lead to
$$
\imath \hbar \frac{\partial \psi}{\partial t} = \frac{-\hbar^2}{2m'}
\frac{\partial^2 \psi}{\partial x^2} + \int \psi^* (x')\psi (x) \psi
(x')U(x')dx', \quad \quad \quad (2)$$

In the above $U(x') = 1 \, \mbox{for} \, x'$ in a $\delta$ interval,
a small interval around this point and $= 0$ outside [1,2].\\
In the Landau-Ginsburg case there is a coherence length which is
given by
$$
\xi = \left(\frac{\gamma}{\alpha}\right)^{\frac{1}{2}} = \frac{h
\nu_F} {\Delta}\quad \quad \quad \quad \quad (3)
$$
which now appears as the Compton wavelength. From the slightly
different analysis of Planck oscillators we come to the same
conclusion [3]. So the picture that emerges is, starting with
Wheeler's Quantum Foam,[4], presumably immediately after the Big
Bang, we are lead to the Compton scale.\\ \\
2. Let us come to the problem of the cosmological constant. This was
noticed some decades ago by Zeldovich and others and become well
known as the cosmological constant problem. The problem is that if
we consider the Zero Point Energy at the Planck scale, the
cosmological constant which is the vacuum energy density becomes
enormous. Roughly give the Planck scale this would be of the order
$$
\frac{mc^2}{l^3} \sim 10^{115}\quad \quad \quad \quad \quad \quad
(4)
$$
This enormous value is some $10^{120}$ times the observed value. But
now let us consider this at the Compton scale. Then as can be seen
from (4) the cosmological constant would be reduced by a
factor of $10^{80}$ aligning it with observation.\\ \\
4. The above argument in fact provides us with a unified description
of electromagnetism and gravitation. It is well known in spite of a
century's effort, starting from Hermann Weyl, right up to string
theory there has been no satisfactory ''unification'' of
electromagnetism and gravitation. In fact Pauli observed that we
should not try to unify what nature had meant to be separate. But
let us consider the following argument: First let us invoke the work
of Cercignani [5] in a pre dark energy era. He used Quantum
oscillations invoking the usual Zero Point Field. He showed, using
the fact that mass and energy were equivalent, that chaotic
oscillations are present whenever mass is of the order
$$
G [\hbar \omega c^{-2}]^2 [\omega^{-1}c]^{-1} = G\hbar^2 \omega^3
c^{-5}\quad \quad \quad \quad (5)
$$
where $G$ is the constant of gravitational attraction and we have
used the wavelength for the distance.\\
If this were to be less than the electromagnetic energy $\hbar
\omega$ then we must have
$$
(G\hbar) ^{-1/2} \cdot c^{5/2}\quad \quad \quad \quad \quad \quad
(6)
$$
This is what may be called a gravitational cut off for the frequency
in the Zero Point Energy. In other words above this cut off
frequency for the Zero Point Energy we have gravitation but below it
we come to the realm of electromagnetism. This maximum frequency
oscillation is given by
$$G\hbar \omega^2_{max} = c^5\quad \quad \quad \quad \quad (7)$$
Interestingly (7) shows that at the Planck scale the electromagnetic
and gravitational strengths are of the same order. However after the
phase transition when we come to the Compton scale, we have the
usual electromagnetic field. So the phase transition leads from
gravitation to electromagnetism, providing a unified description.\\
\\
\begin{flushleft}
{\large {\bf References for the Appendix}}
\end{flushleft}
\noindent [1] Sidharth, B.G. (1997). \emph{Ind. J. of Pure and Applied Physics} 35, pp.456ff.\\
\noindent [2] Sidharth, B.G. (2005). \emph{The Universe of
Fluctuations} (Springer, Netherlands).\\
\noindent [3] Sidharth, B.G. (2008). \emph{Thermodynamic Universe}
(World Scientific, Singapore).\\
\noindent [4] Wheeler, J.A. and Kenneth, F. (1993). \emph{Geons,
Black Holes and Quantum Foam}, (W.W. Norton \& Co. New York, 1993).\\
\noindent [5] Cercignani, C. (1998). \emph{Found.Phys.Lett.} Vol.11,
No.2, pp.189-199.
\end{document}